%% file: arxive.tex
\pdfoutput=1

\documentclass{article}% uncomment this for twocolumn layout and comment line below
\usepackage{url}
\usepackage{algpseudocode}
\usepackage{amsmath}
\usepackage{colortbl}

\usepackage{geometry}
\usepackage{amsmath, amsthm, amssymb}
\usepackage{algorithm,graphicx}
\usepackage{algpseudocode}
\usepackage{fixltx2e} 
\usepackage{amsmath}
\usepackage{sidecap}

\def\MSL{\mbox{\rm {\sf MSL}}}
\def\SP{\mbox{\rm {\sf SP}}}
\def\SA{\mbox{\rm {\sf SA}}}
\def\A{\mbox{\rm {\sf A}}}
\def\I{\mbox{\rm {\sf I}}}
\def\LCP{\mbox{\rm {\sf LCP}}}

\newtheorem{theorem}{Theorem}
\newtheorem{corollary}[theorem]{Corollary}

\newtheorem{definition}[theorem]{Definition}

\begin{document}

%%% Start of article front matter
%\begin{frontmatter}

%\begin{fmbox}
%\dochead{Research}

%%%%%%%%%%%%%%%%%%%%%%%%%%%%%%%%%%%%%%%%%%%%%%
%%                                          %%
%% Enter the title of your article here     %%
%%                                          %%
%%%%%%%%%%%%%%%%%%%%%%%%%%%%%%%%%%%%%%%%%%%%%%
\title{HyDA-Vista: Towards Optimal Guided Selection of $k$-mer Size for Sequence Assembly}
\author{Seyed Basir Shariat Razavi*\\ Narjes Sadat Movahedi Tabrizi**\\ Hamidreza Chitsaz*\\Christina Boucher*\\ \\
* Department of Computer Science, Colorado State University\\ Fort Collins, CO, USA, \{basir,cboucher,chitsaz\}@cs.colostate.edu\\ 
** Department of Computer Science, Wayne State University\\ Detroit, MI, USA,  narges@wayne.edu\\ }
\maketitle

\input{abstract}
%%%%%%%%%%%%%%%%%%%%%%%%%%%%%%%%%%%%%%%%%%%%%%
%%                                          %%
%% The keywords begin here                  %%
%%                                          %%
%% Put each keyword in separate \kwd{}.     %%
%%                                          %%
%%%%%%%%%%%%%%%%%%%%%%%%%%%%%%%%%%%%%%%%%%%%%%

%\begin{keyword} 
%\kwd{Genome Assembly} 
%\kwd{Sequence Landscape}
%\end{keyword}
%\end{abstractbox}
%
%\end{fmbox}% uncomment this for twcolumn layout

%\end{frontmatter}

%%%%%%%%%%%%%%%%%%%%%%%%%%%%%%%%%%%%%%%%%%%%%%
%%                                          %%
%% The Main Body begins here                %%
%%                                          %%
%% Please refer to the instructions for     %%
%% authors on:                              %%
%% http://www.biomedcentral.com/info/authors%%
%% and include the section headings         %%
%% accordingly for your article type.       %%
%%                                          %%
%% See the Results and Discussion section   %%
%% for details on how to create sub-sections%%
%%                                          %%
%% use \cite{...} to cite references        %%
%%  \cite{koon} and                         %%
%%  \cite{oreg,khar,zvai,xjon,schn,pond}    %%
%%  \nocite{smith,marg,hunn,advi,koha,mouse}%%
%%                                          %%
%%%%%%%%%%%%%%%%%%%%%%%%%%%%%%%%%%%%%%%%%%%%%%

%%%%%%%%%%%%%%%%%%%%%%%%% start of article main body
% <put your article body there>

%%%%%%%%%%%%%%%%
%% Background %%
%%

\input{introduction}
\input{background}

\input{seq_landscape}
\input{approach_landscape}
\input{approach_readassignment}
\input{approach_assembly}
\input{results}
\input{conclusion}

\newpage
%\input{supplement}

%%%%%%%%%%%%%%%%%%%%%%%%%%%%%%%%%%%%%%%%%%%%%%
%%                                          %%
%% Backmatter begins here                   %%
%%                                          %%
%%%%%%%%%%%%%%%%%%%%%%%%%%%%%%%%%%%%%%%%%%%%%%

%\begin{backmatter}

%\section*{Competing interests}
%  The authors declare that they have no competing interests.

%\section*{Author's contributions}
%    Text for this section \ldots

%\section*{Acknowledgements}
  %Text for this section \ldots
%%%%%%%%%%%%%%%%%%%%%%%%%%%%%%%%%%%%%%%%%%%%%%%%%%%%%%%%%%%%%
%%                  The Bibliography                       %%
%%                                                         %%
%%  Bmc_mathpys.bst  will be used to                       %%
%%  create a .BBL file for submission.                     %%
%%  After submission of the .TEX file,                     %%
%%  you will be prompted to submit your .BBL file.         %%
%%                                                         %%
%%                                                         %%
%%  Note that the displayed Bibliography will not          %%
%%  necessarily be rendered by Latex exactly as specified  %%
%%  in the online Instructions for Authors.                %%
%%                                                         %%
%%%%%%%%%%%%%%%%%%%%%%%%%%%%%%%%%%%%%%%%%%%%%%%%%%%%%%%%%%%%%

% if your bibliography is in bibtex format, use those commands:
\bibliographystyle{plain} % Style BST file
\bibliography{document}
%\input{supplement}

%\end{backmatter}
\end{document}

%% file: abstract.tex
\begin{abstract}
{Motivation:} Intimately tied to assembly quality is the complexity of the de Bruijn graph built by the assembler.  Thus, there have been many paradigms developed to decrease the complexity of the de Bruijn graph.  One obvious combinatorial paradigm for this is to allow the value of $k$ to vary; having a larger value of $k$ where the graph is more complex and a smaller value of $k$ where the graph would likely contain fewer spurious edges and vertices.  One open problem that affects the practicality of this method is how to predict the value of $k$ prior to building the de Bruijn graph.  We show that optimal values of $k$ can be predicted prior to assembly by using the information contained in a phylogenetically-close genome and therefore, help make the use of multiple values of $k$ practical for genome assembly.  
 
{Results:}
We present HyDA-Vista, which is a genome assembler that uses homology information to choose a value of $k$ for each read prior to the de Bruijn graph construction. The chosen $k$ is optimal if there are no sequencing errors and the coverage is sufficient. Fundamental to our method is the construction of the {\em maximal sequence landscape}, which is a data structure that stores for each position in the input string, the largest repeated substring containing that position. In particular, we show the maximal sequence landscape can be constructed in $O(n + n \log n)$-time and $O(n)$-space. HyDA-Vista first constructs the maximal sequence landscape for a homologous genome. The reads are then aligned to this reference genome, and values of $k$ are assigned to each read using the maximal sequence landscape and the alignments. Eventually, all the reads are assembled by an iterative de Bruijn graph construction method. Our results and comparison to other assemblers demonstrate that HyDA-Vista achieves the best assembly of {\em E. coli} 
%and \emph{Arabidopsis} chromosome four
 before repeat resolution or scaffolding. 

{Availability:}
HyDA-Vista is freely available at \url{https://sites.google.com/site/hydavista}.  The code for constructing the maximal sequence landscape and the choosing the optimal value of $k$ for each read is also on the website and could be incorporated into any genome assembler.

{Contact:} {basir@cs.colostate.edu}
\end{abstract}

%\end{abstractbox}

%% file: introduction.tex
\section{Introduction}

The ability to accurately assemble genomes is a fundamental problem in bioinformatics that is vital to the success of many scientific projects, including the 10,000 vertebrate genomes (Genome 10K) \cite{Haussler09},  \emph{Arabidopsis} variations (1001 genomes) \cite{Ossowski08}, human variations (1000 genomes) \cite{Abecasis12}, and Human Microbiome Project \cite{hmp}.  The genome assembly process aims to build contiguous sequences, called {\em contigs}, predominantly from short read sequencing data. Other sources of information have also been used to boost the accuracy, including genetic linkage data \cite{lin2012agora}, optical mapping data \cite{nagarajan2008scaffolding}, and longer sequencing reads (e.g. PacBio data) \cite{huddleston2014}.  A potential source of information that has not been fully explored is the information contained in phylogenetically-close genomes. The genome of an individual of the same species or that of a phylogenetically-close species can potentially be used as an extra source of information, and increase the assembly quality.  We argue that genome assemblers can benefit from using a reference genome to help guide the assembly process, particularly in those regions of the genome that are pervaded by repetitive sequences. 

In Eulerian sequence assembly \cite{IW95,PTW}, a de Bruijn graph is constructed with a vertex $v$ for every $(k - 1)$-mer present in an input set of reads, and an edge $(v, v')$ for every observed $k$-mer in the reads with $(k - 1)$-mer prefix $v$ and $(k - 1)$-mer suffix $v'$. A contig corresponds to a non-branching path through this graph.  SPAdes \cite{bankevich2012spades}, IDBA \cite{peng2010idba}, Euler-SR \cite{Chaisson:2008}, Velvet \cite{Zerbino:2008}, SOAPdenovo \cite{Li:2010}, ABySS \cite{Simpson:2009} and ALLPATHS \cite{Butler:2008} all use this paradigm for assembly. The majority of de Bruijn graph based assemblers follow the same general outline: break the (possibly error corrected) reads into $k$-mers, construct the de Bruijn graph on the set of resulting $k$-mers, simplify the de Bruijn graph, resolve the repeated regions by using mate-pair information, and construct the contigs (simple paths in the de Bruijn graph). Therefore, the majority of assemblers require or allow the value of $k$ to be specified by the user.

The problem of determining an appropriate value of $k$ for the de Bruijn graph construction is important since it has a direct impact on assembly quality; stated very briefly, when $k$ is too small the resulting graph is complicated by spurious edges and vertices, and when $k$ is too large the graph becomes too sparse and possibly disconnected.  Repetitive regions are especially problematic for genome assembly since they inadvertently result in spurious edges and vertices in the de Bruijn graph \cite{alkan2011limitations} and are very sensitive to the choice of $k$.  There has been a significant effort in developing algorithms that will choose an ideal value for $k$ by preprocessing the sequence reads, and thus, reduce the complexity of the de Bruijn graph \cite{bankevich2012spades,peng2010idba,kmergenie13}. 

A more obvious combinatorial approach for building a simplified de Bruijn graph would be to allow the value of $k$ to vary; having a larger value of $k$ where the graph is more complex and a smaller value of $k$ where the graph would likely contain fewer spurious edges and vertices.  A major difficulty in implementing this approach is determining a practical method that makes this idea feasible assembling large genomes.  Peng \emph{et al.}~\cite{peng2010idba} and Bankevich \emph{et al.}~\cite{bankevich2012spades} both introduced assemblers that use various values of $k$. IDBA builds the de Bruijn graph in an iterative manner from $k=k_{min}$ to $k=k_{max}$; these values of $k$ are predetermined and (by default) do not change for different datasets or genomes. At iteration $i$, the de Bruin graph for $k_i$ is built from the current set of reads and the contigs for that graph are constructed, then all the reads that align to at least one of those contigs are removed from the current set of reads. In the next iteration the graph is built by converting every edge from the previous graph to a vertex while treating contigs as edges. SPAdes \cite{bankevich2012spades} uses a similar approach but uses all the reads at each iteration.  

While this method has been shown to greatly improve assembly quality \cite{bankevich2012spades,peng2010idba}, it is not efficient since all the reads are assembled at each iteration.  Thus, one challenge that remains to be addressed is how to efficiently determine which values of $k$ should be used for this iterative assembly process and how to assign a $k$-mer value for each read.  If this could be accomplished prior to assembly of the de Bruijn graph(s) then these iterative assembly methods could be made more efficient without degrading the assemblies quality.

\paragraph{Our Contribution.} We introduce an efficient algorithm for determining an optimal value of $k$ for each read prior to constructing the de Bruijn graph, and implement this method into a modified version of HyDA, a {\em de novo} assembler developed by Movahedi \emph{et al.} \cite{hyda}. This modified assembler, which we refer to as \emph{HyDA-Vista}, takes as input a phylogenetically-close genome and a set of paired-end reads.  Imperative to HyDA-Vista is the construction of the {\em maximal sequence landscape}, which is a data structure that stores for every position in the input string, the longest repeat containing it. Prior to de Bruijn graph construction, HyDA-Vista constructs the maximal sequence landscape for the  phylogenetically-close genome, and aligns the reads to the reference genome.  The alignment and landscape allows the optimal value of $k$ for each read be determined in linear time in the length of the read, provided the read is longer than the longest repeat.  These values of $k$ are ``optimal'' in the sense that for unchanged parts of the genome, the de Bruijn graph will have no spurious edges or vertices if there are no sequencing errors, and the length of the repeat is smaller than the read length.  Unaligned reads are assigned a default value of $k$.  After the assignment of values of $k$ to each read, HyDA-Vista constructs the de Bruijn graph in an iterative manner.  

Our approach for choosing values of $k$ for each read takes into consideration the repeat structure of the genome, which enables us to avoid overly-complex regions of the graph since the assignment of values of $k$ to reads is done prior to the assembly rather than during the assembly.   We compare HyDA-Vista versus IDBA \cite{peng2010idba}, SPAdes \cite{bankevich2012spades}, SOAPdenovo \cite{Li:2010}, ABySS \cite{Simpson:2009} and HyDA \cite{hyda}.  Our results demonstrate that HyDA-Vista produces the best assembly of {\em E. coli} before repeat resolution or scaffolding.  We aim to achieve the best assembly without repeat resolution and scaffolding and note that such methods could be applied to all these initial assemblies.  Lastly, we demonstrate that this method improves the efficiency of iterative assembly.

%IDBA-UD \cite{idbaud},

%Hence, a unique feature of our method is that the de Bruijn graph is built in an informed manner for major portions of the genome.  

\paragraph{Roadmap.} We review related tools for the problem in the remainder of this section. Section~\ref{sec:background}  then sets notation and formally lays down the problem and the data structures that will be used for the construction of the maximal sequence.  We formally define the maximal sequence landscape in Section~\ref{sec:landscape}.  Section~\ref{sec:graph} gives details of HyDA-Vista.  In Section~\ref{sec:results} we give results that demonstrate how HyDA-Vista compares against competing assemblers. Finally, Section~\ref{sec:conclusion} offers reflections and some future areas of research that warrant investigation.

\paragraph{Related Work.} The re-sequencing methods include LOCAS and SUPERLOCAS \cite{schneeberger2011reference,klein2011locas}, e-RGA \cite{Vezzi2011}, Colombus module of Velvet \footnote{Unpublished. \url{http://bioweb2.pasteur.fr/docs/velvet/Columbus_manual.pdf}} and IDBA-Hybrid \footnote{Unpublished. \url{http://i.cs.hku.hk/~alse/hkubrg/projects/idba_hybrid/index.html}}. Since these methods aim to determine structural variations between species, and require extremely high sequence similarity to produce reasonable results, they have only been applied to individuals of the same species. Our focus is to produce high quality {\em de novo} assemblies using homology information contained in the reference genome of the same species or phylogenetically-close species.    Gnerre et al.~\cite{gnerre2009assisted} also consider how to improve assembly quality by using the alignment of reads to a reference genome.  Their method simultaneously builds a {\em de novo} assembly from the reads and aligns these same reads to one or more related genomes. The alignment is then used to improve the assembly quality, e.g., reads that were not used in the assembly are incorporated into the assembly using the alignment. 
%For example, the whole-genome of four cultivars of {\em Arabidopsis} were assembled using a method that is based on extracting variants from a reference sequence using sequence alignment 

Complementary to the work of Chikhi and Medvedev \cite{kmergenie13}, Peng \emph{et al.}~\cite{peng2010idba}, and Bankevich \emph{et al.}~\cite{bankevich2012spades}, there has been an effort in developing methods that use paired-end data to constrain the construction of the de Bruijn graph \cite{schneeberger2011reference,pevzner2004novo,phillippy2008genome,medvedev2011paired,recGraphs}.
Medvedev et al.~\cite{medvedev2011paired} introduced the concept of a paired de Brujin graph
% where the vertices represent pairs of $k$-mers in mate pairs, and there is an edge between two vertices when their corresponding bi-mers satisfy the insert size (distance) constraints
. Since the insert size is variable among mate pairs, this method requires that all the paths within some threshold be considered in order to ensure an edge is not missed.  Thus, Bankevich \emph{et al.} \cite{bankevich2012spades} improve upon this idea by developing the rectangle graph, which eliminates the need to consider all paths.  Vyahhi \emph{et al.}~\cite{recGraphs} furthered this study of  rectangle graphs for genome assembly.  These methods merit mentioning the goal of these methods is the same as the goal of  increasing the value of $k$ in certain regions; both aim to minimize spurious edges and branching in the graph but in a different manner.

Determining all maximal exact repeats in a string has been previously studied \cite{gusfield1997algorithms,mcconnell86,Blumer198531}. It has been shown that all maximal repeats of a string can be found and stored in $O(n)$-time and $O(n)$-space using a suffix tree (although the output maybe of size $\Theta(n^2)$) or directed acrylic graph \cite{mcconnell86}. Therefore, the maximal sequence landscape, which we define in this paper, can be constructed from either a suffix tree or a directed acrylic graph in $O(n)$-time and $O(n)$-space using these algorithms directly or adapting them.  However, the constant in the order notation of the space complexity of these constructions is relatively large. The algorithm we present uses a suffix array and thus, requires linear space with a smaller constant and $O(n\log n)$-time.  Hence, we pay a $\log n$ cost in time to remove the large constant from the linear space time. We also note that the related problem of finding inexact maximal repeats also has been previously studied \cite{kurtz2001reputer,fitch1985detecting,benson1995space,sagot1998spelling}.

%% file: background.tex
\section{Background} \label{sec:background}

\paragraph{Strings.}
Consider a fixed alphabet $\Sigma = \{\sigma_1,\ldots,\sigma_m\}$ and a total order $\leq_L$ defined over $\Sigma = \Sigma \cup \{\$\}$ where $\$ \notin \Sigma$, and for all $\sigma \in \Sigma$ we have $\$ \leq_{L} \sigma$. We denote a finite string $s$ as $s_1s_2 \ldots s_n$, where $s_i \in \Sigma$.  % \emph{etc.}\footnote{We assume that the size of the alphabet is a fixed constant through out this paper ($m \in O(C)$).}. 
We use $s_{ij}$, where $1\leq i\leq j \leq n$, to indicate substring $s_is_{i+1}\cdots s_{j}$ of string $s$. 
We call substrings $s^{\mbox{pre}}_i=s_{1i}$ and $s^{\mbox{suf}}_i=s_{in}$ with $i\in\{1,\ldots,n\}$ the $i$\textsuperscript{th} \emph{prefix} and $i$\textsuperscript{th} \emph{suffix} of $s$ respectively. Based on the total order $\leq_L$, we define a \emph{lexicographical} total order on the strings in $\Sigma^*$. %Wrong definition: For all $s,t \in S$, we say that string $s=s_1\cdotss_n\$$ is \emph{lexicographically smaller than} string $t=t_1\cdots t_r\$$ and write $s \leq_L t$ if and only if $s_i \leq_L t_i$ for all $i\in \{1, \ldots,\min(n,m)+1 \}$. 

\paragraph{Arrays.}
We denote arrays of integers by all capital letter strings like $\A$, $\SP$, $\LCP$, etc.  $\A[i]$, with $1 \leq i\leq |\A|$, stands for the integer in the  $i$\textsuperscript{th} cell of array $\A$. Also, $\A[i,j]$ indicates the projection of $\A$ onto indices $i$ to $j$, inclusive of both ends. For an array $\A$, with $|\A|=n$, that holds a permutation of integers $\{1, \ldots, n\}$, \emph{index array} of $\A$ is another array $\I(\A)$ with $| \I(\A)|=n$ such that $\I(\A)[i]=j$ if and only if $\A[j]=i$.

\paragraph{Suffix and Longest Common Prefix Arrays.}
$\SA_s$, for some string $s$, denotes the \emph{suffix array} associated with $s$ \cite{mm1993}. $\SA_s[i]=l$ for $i \in \{1, \ldots, n\}$ if and only if $s^{\mbox{suf}}_l$ is the $i$\textsuperscript{th} string in the lexicographically sorted list of all suffixes of $s$. We also indicate the \emph{longest common prefix array} of some string $s$ with $\LCP_s$, and 
$\LCP_s[i]=l$ for $i \in \{1,\ldots, n-1\}$ if and only if the length of the longest common prefix between $s^{\mbox{suf}}_{\SA_s[i]}$ and $s^{\mbox{suf}}_{\SA_s[i+1]}$ equals to $l$. 

%\noindent
%\textbf{Array Intervals:}
%For strings $s=s_1..s_n$ and $q = q_1..q_l$ and integer $i \in \{1,..,l\}$, $SA_s[q^p_i]$ is the interval $[a,b]$ with $1\leq  a \leq b \leq n$ such that for all $j \in [a,b]$, $q^p_i$ is a prefix of $SA_s[j]$. Since $SA_s[q^p_i]$ is an interval, one can safely write $LCP_s[SA_s[q^p_i]]$ which is the limitation of array $LCP_s$ to the indices in interval $SA_s[q^p_i]$.

%% file: seq_landscape.tex
\section{Approach}

\begin{figure}
\centering
\includegraphics[scale=0.4]{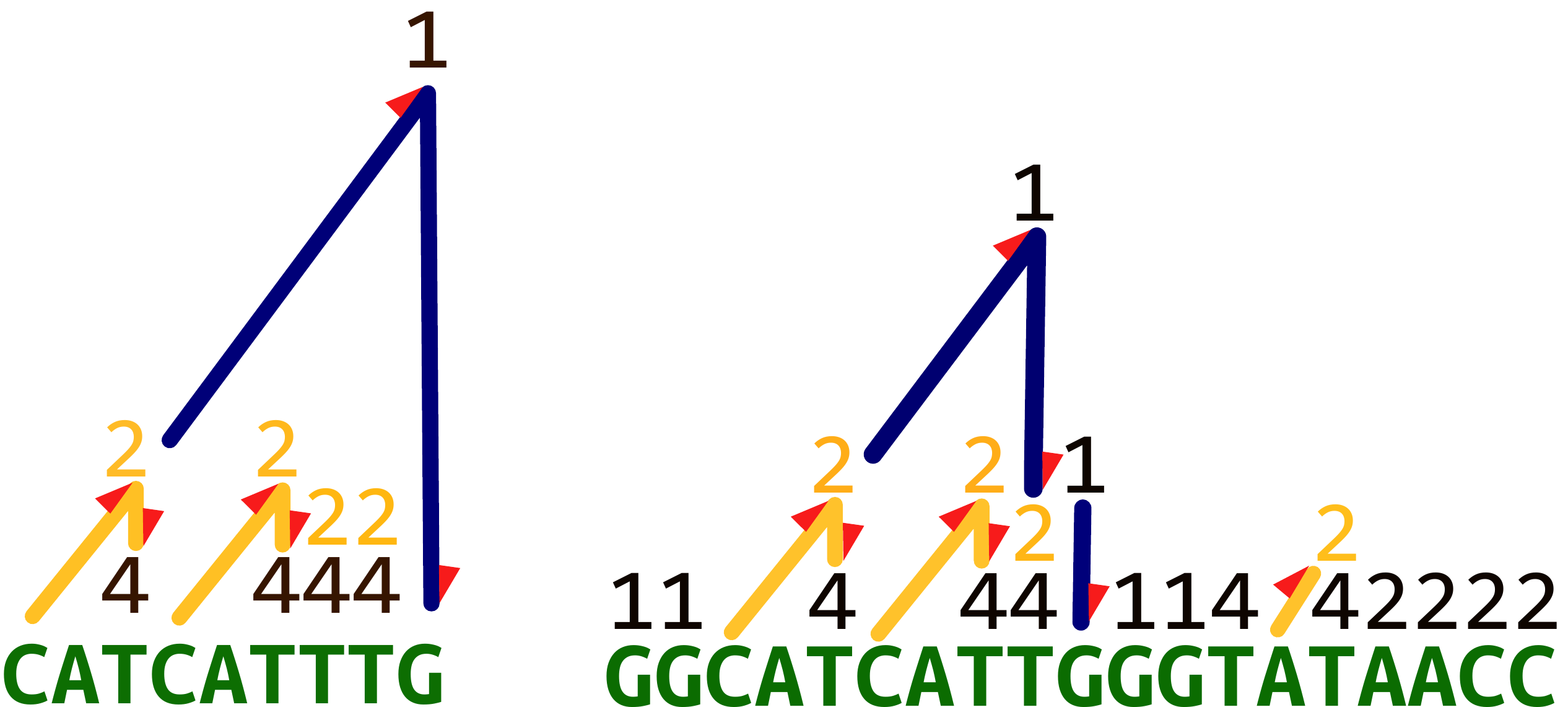}
\caption{(left) The self sequence landscape for CATCATTTG, and (right) the sequence landscape of GGCATCATTGGGTATAACC with respect to CATCATTTG. The mountains (yellow or blue) demonstrate occurrences of substrings of source string in the target string. Numbers at the peak of the mountains denote the frequency of occurrence. The maximal sequence landscape is highlighted in yellow, and the red arrows demonstrate the ascent and descent of the landscapes.}\label{fig:land}
\end{figure}

\subsection{The Maximal  Sequence Landscape} \label{sec:landscape}

We formally define the maximal sequence landscape in this section.  Clift \emph{et al.} \cite{mcconnell86} introduced the concept of a sequence landscape, which is a data structure that stores the occurrences of any substring from a source string $s$ in a target string $t$.  In set representation, the sequence landscape $L_{t|s}$ of a target string $t$ with respect to a source string $s$ is defined as a set of tuples $\{m_1, m_2, \ldots, m_l\}$, where $m_i = (b_i, e_i, f_i)$ corresponds to the occurrence of substring $s_{b_ie_i}=s_{b_i}s_{b_i+1}\cdots s_{e_i}$ from $s$ in  $t$ with frequency $f_i$. If $s$ and $t$ are equal then the sequence landscape categorizes all repeated substrings in the source string $s$.  We define to this special case where $s = t$ as the {\em self sequence landscape}.  Fig. \ref{fig:land} illustrates an example of a self sequence landscape and a sequence landscape.  Given a position $i$ of the input string $s$, all the repeated substrings containing $s_i$ can be recovered from the self sequence landscape in linear-time in the number of different  repetitions.  %For example, the self sequence landscape for the input string GTAGTAGTAGTA will store the repetitions for GTA, GTAGTA for 12$th$ position.  

The occurrences of the substrings in the source string are defined as {\em mountains}.  This terminology reflects the visual representation that was first introduced by Clift \emph{et al.} \cite{mcconnell86} that illustrates each occurrence as a mountain having height equal to the length of the substring, i.e. the height of mountain $m_i$ of $L_{t|s}$ is denoted as $h(m_i)$ and equal to $e_i - b_i + 1$. The {\em peak} of each mountain is labelled with the frequency of the substring corresponding to it.  In Fig. \ref{fig:land} (left), the substring CAT is represented as two mountains each of which has a height equal to three and frequency equal to two.

We say that a \emph{mountain} $m_j = (b_j, e_j, f_j)$ in a landscape $L_{t|s}$ \emph{covers} index $i$ and denote it by $m_j \bigtriangleup i$ if and only if $i \in \{b_j, \ldots, e_j\}$. Hence, the \emph{cover set} of a specific index $i$ of the sequence landscape $L_{t|s}$ is the set of all the mountains that covers $i$.  We denote the cover set as $C_{L_{t|s}}(i)$ and define it as follows:
\begin{equation}
C_{L_{t|s}}(i) := \{m_j|m_j \bigtriangleup i,h(m_j)>1,f_j>1\}.
\end{equation}
Lastly, we define the \emph{summit of index} $i$ as the highest mountains in its cover set.   We denote the summit of $i$ by $S_{L_{t|s}}(i)$ and define it as follows: 
\begin{equation}
S_{L_{t|s}}(i) :=  \{ m_j \, | \,h(m_j) \geq h(m_k) \,\, \forall \, m_k \in C_{L_{t|s}}(i)\}.
\end{equation}

Please note that the summit of index $i$ can be empty or non-unique so the height of summit of index $i$ is defined to be zero for empty set. 

\begin{definition} The \emph{maximal sequence landscape}, which we denote as $L^*_{t|s}$, is the set of the summits of all positions in $s$ that have frequency greater than one.  $L^*_{t|s}$ can be formally defined as follows: $L^*_{t|s} = \{ S_{L_{t|s}}(i) | i=1,\ldots, n\}$. The maximal sequence landscape is highlighted in yellow in Fig. \ref{fig:land}. \end{definition}

The maximal sequence landscape is obtained from the sequence landscape by removing all mountains except those that are highest and have frequency greater than one at each position.  In the case of the maximal sequence landscape constructed from a self sequence landscape, this results in a data structure containing the longest repeat at each position of the input string.  In Subsection \ref{construction} we give an algorithm that builds the maximal sequence landscape and returns an array containing the length of the longest repeat at each position of the input string ($\MSL$ in Algorithm \ref{alg:landscape}).  Therefore, given a position $i$ in $s$, we can determine the length of the longest repeat in $s$ containing that position in constant time by simply indexing the maximal sequence landscape at position $i$. By choosing a value for $k$ that is larger than the length of this repeat it can guaranteed that there will be no branching in the corresponding vertices of the de Bruijn graph, if the same substring is not repeated in changed parts of the genome that is being assembled. This is our idea based on which we determine the optimal value of $k$ for each read.  We consider the maximal sequence landscape constructed from the self sequence landscape for the remainder of this paper since it is what is used by HyDA-Vista.

%% file: approach_landscape.tex
\section{Methods}

Algorithm \ref{alg:hyda} gives an overview of HyDA-Vista algorithm.   We explain each of these steps in detail in the subsequent subsections. %The MCL-WMR algoforithm first chooses a {\em reference sequence} from the data set, denoted as $S_r$, building the entire graph from all the input data (including $S_r$), and for each vertex $v_{r \gamma}$ representing the $l$-length subsequence from $S_r$ starting at position $\gamma$.  We then use MCL to generate subgraphs which contain vertices that are highly inter-related, from these clusters of vertices we will generate the positions of the possible motif instances and their corresponding motif consensus. The algorithm terminates when a motif is found.  In order to increase the probability a motif is found, we minimize searching subgraphs with low probability of containing a motif; hence, the adjacency subgraphs are not clustered and searched in a sequential manner.

\begin{algorithm}[h!]
\caption{An overview of HyDA-Vista}
\label{alg:hyda}
\begin{algorithmic}[1]
\State  Build the maximal sequence landscape for the reference genome.
\State  Align all reads to the reference using BWA.
\State  For each aligned read: assign a value of $k$ using the maximal sequence landscape.
\State  Unaligned reads are assigned a value of $k$ using a heuristic.
\State  The de Bruijn graph is constructed in an iterative manner, as shown in Algorithm \ref{alg:assembly}.
\end{algorithmic}
\end{algorithm}

\begin{algorithm*}
\caption{Maximal sequence landscape construction}
\label{alg:landscape}
\begin{algorithmic}[1]
\Require String $s$ of length $n$.
\Ensure Maximum Sequence Landscape Array $\MSL$.

\State  $\MSL \gets \emptyset$
\State  Build $\SA_s$, $\LCP_s$, $\SP_s$, and the ordered binary search tree of the $\LCP_s$
\State  $b \gets 0, e \gets 0$ \Comment {$b$ and $e$ denote positions in the string $s$}
\State  $[sa_{\min}, sa_{\max}] \gets [1, n]$ %\Comment {$[min_i,max_i]$ denotes the search interval}
\For {$i:=1$ {\bf to} $n$}
	\State	$e \gets e + 1$, $p \gets s_b s_{b+1} \cdots s_e$ 
	\If{$p \in \SA_s$ and has frequency greater than one}
		\State  Update $[sa_{\min}, sa_{\max}]$ \Comment {A repeat has been found}
	\Else 
		\While{$b \neq e $ and no repeat is found}
			\State $b \gets b + 1$, $p = s_b s_{b+1} \cdots s_e$  
			\State Find the new interval $[sa_{\min}, sa_{\max}]$ \Comment {See Section 	\ref{construction}}
%\State  \hspace{15mm}Set $l = SP_s(l)$. \Comment{Find $p$ in $SA_s$ using $SP_s$}
%\State  \hspace{15mm}Set $[min_i,max_i] =$ largest interval around $l$ that for all $m \in [min_i,max_i] $ , $LCP_s[m]\leq |p|-1$
			\If{$p \in \SA_s$ and has frequency greater than one}  
				\State Update $[sa_{\min}, sa_{\max}]$ \Comment {A repeat has been found}
			\EndIf
		\EndWhile
	\EndIf
	\State Update $\MSL[i]$  %\gets b-e+1$
\EndFor
\end{algorithmic}
\end{algorithm*}

\begin{table*}[t]
\begin{small}

\centering
\caption{Construction of the suffix pointer array ($\SP_s$) using the suffix array. The dark column in each table denotes the index of the array. We build an inverse index $I(\SA_s)$ from the suffix array, and $\SP_s$ can be constructed by scanning this array once, i.e. setting $\SP_s[I(\SA_s)[i-1]]=I(\SA_s)[i]$ for all $i \in [2, n]$ and $\SP_s[1] = 0$.}
\label{fig:sp}
\begin{tabular}{|>{\columncolor[gray]{0.5}}c|c|>{\columncolor[gray]{0.8}}l|}
\hline
\multicolumn{3}{|c|}{Suffix Start}\\ \hline
1 &1 & mybananas\$ \\\hline
2&2& ybananas\$ \\ \hline
3&3& bananas\$ \\ \hline
4&4& ananas\$ \\ \hline
5&5& nanas\$ \\ \hline
6&6& anas\$\\ \hline
7&7& nas\$ \\ \hline
8&8& as\$ \\ \hline
9&9& s\$ \\ \hline
10&10& \$ \\ \hline
\end{tabular}
$\Longrightarrow$
\begin{tabular}{|>{\columncolor[gray]{0.5}}c|c|c|>{\columncolor[gray]{0.8}}l|}
\hline
\multicolumn{4}{|c|}{Suffix \& \LCP Arrays}\\ \hline
1&10&0& \$ \\ \hline
2&4&0& ananas\$ \\ \hline
3&6&3& anas\$\\ \hline
4&8&1& as\$ \\ \hline
5&3&0& bananas\$ \\ \hline
6&1&0& mybananas\$ \\\hline
7&5&0& nanas\$ \\ \hline
8&7&2& nas\$ \\ \hline
9&9&0& s\$ \\ \hline
10&2&0& ybananas\$ \\ \hline
\end{tabular}
$\Longrightarrow$
\begin{tabular}{|>{\columncolor[gray]{0.5}}c|c|}
\hline
\multicolumn{2}{|c|}{SA Index}\\ \hline
1&6  \\ \hline
2&10 \\ \hline
3&5  \\ \hline
4&2  \\ \hline
5&7  \\ \hline
6&3  \\ \hline
7&8  \\ \hline
8&4  \\ \hline
9&9  \\ \hline
10&1 \\ \hline
\end{tabular}
$\Longrightarrow$
\begin{tabular}{|>{\columncolor[gray]{0.5}}c|c|}
\hline
\multicolumn{2}{|c|}{\SP}\\ \hline
1&0 \\ \hline
2&7 \\ \hline
3&8 \\ \hline
4&9 \\ \hline
5&2 \\ \hline
6&10 \\ \hline
7&3 \\ \hline
8&4 \\ \hline
9&1 \\ \hline
10&5 \\ \hline
\end{tabular}

\end{small}
\end{table*} 

\subsection{Construction of the Maximal Sequence Landscape} \label{construction}

%Our algorithm relies on the use of suffix and longest common prefix arrays.

In this section we demonstrate that the maximal sequence landscape for an input string $s$ can be built in $O(n + n \log n)$-time and $O(n)$-space using a very simple algorithm, where $n$ is the length of $s$. Algorithm \ref{alg:landscape} gives the pseudocode for constructing the maximal sequence landscape.   Our method relies on the use of suffix array and longest common prefix array and thus, begins by building the suffix array ($\SA_s$), and the longest common prefix array ($\LCP_s$).  This construction can be done in $O(n)$-space and $O(n)$-time \cite{ks2006}.  Two other auxiliary data structures are constructed at the beginning of the algorithm. However, we delay the definition and construction of those to later in this section.  The algorithm then iterates through each position of $s$ and finds the longest repeated substring in $s$ that contains it using $\SA_{s}$, $\LCP_{s}$, and the auxiliary data structures.  An important aspect of our algorithm that allows us to achieve 
$O(n + n \log n)$-time is that we only search that interval of $\SA_s$ which is between the indices $\SA_{\min}$ and $\SA_{\max}$ at each iteration, not the entire array. In other words this invariant holds at each iteration of our algorithm: $[\SA_{\min},\SA_{\max}]$ holds the the interval in the suffix array that corresponding suffixes share a same prefix. This prefix is the longest repeat that has been seen so far and covers that position.  Thus, each time the largest repeated substring is found for a particular position, the maximal sequence landscape (
$\MSL$), $\SA_{\min}$, and $\SA_{\max}$ are updated for search at the next iteration.  Two possibilities exist at each iteration $i$ of the algorithm when we are processing $s_i$;

\begin{enumerate}
\item[(a)]
The longest repeated substring at position $i-1$ can be extended by appending $s_i$. The maximal sequence landscape, $\SA_{\min}$, and $\SA_{\max}$ are updated.  %The frequency of occurrence of the new substring remains greater than one 
\item[(b)] The longest repeated substring at position $i-1$ {\em cannot} be extended by appending $s_i$ (either the extended string does not occur or it does occur but its frequency is one). Let $p=s_j \cdots s_{i-1}$ be the longest repeated substring yet found that contains $s_{i - 1}$, and  $p' = s_{j + 1}\cdots s_{i}$ be the string obtained by removing the first letter of $p$ and appending $s_i$. If $p'$'s frequency of occurrence is greater than one, then the maximal sequence landscape, and the search interval is updated as in (a). Otherwise, the search for the longest repeated substring continues by eliminating the first character of $p'$ each time until a repeating match is found or the null string is reached.  If the null string is reached then the maximal sequence landscape is empty at that position and the search interval is updated to $[1, n]$.
\end{enumerate}

The search interval contains all indices in $\SA_s$ for which the corresponding suffixes have the current longest repeated substring as a prefix.  In (a), the interval is updated by performing binary search.  In (b) the search interval is no longer valid since we removed a letter from the \emph{beginning} of the current longest repeated substring and we need (a more complicated) scheme to efficiently find the correct search interval. To accomplish this we need two auxiliary data structures that are constructed at the beginning of the algorithm: the $\SP_s$ array, and an ordered binary search tree containing all consecutive intervals of $\LCP_s$.  $\SP_s[j]$ holds the index in $\SA_s$ that is obtained by removing the first letter from the beginning of $s_{j}s_{j + 1} \cdots s_n$ in order to obtain $s_{j + 1} \cdots s_n$. This array can be built in linear time by scanning the index array of $\SA_s$ (see background for definition of index array). Table \ref{fig:sp} illustrates the construction of $\SP_s$ in linear time.   Thus, to find the correct interval in (b), we locate an index of $\SA_s$ (denoted as $sp$)  where the corresponding suffix contains $s_{j+1} \ldots s_{i-1}$ as a prefix, and find the largest interval around $sp$ where all the suffixes in the interval have $s_{j+1} \ldots s_{i-1}$ as prefix.  This is the new search interval.   The $sp$ index can be found in constant time by correctly indexing $\SP_s$ (see the appendix).  The second step is equivalent to finding the largest interval $[d, u]$ around $sp$ that for all $j \in [d, u]$ we have $\LCP_s[j] \geq |p'| - 1$.  Corollary \ref{lemma2} shows that this can be done in $O(\log n)$-time and $O(n)$-space using an ordered binary search tree.   We leave the proof of the corollary, and the construction of the ordered binary search tree to the appendix.  

\begin{corollary} \label{lemma2}Given a string $p$, $\SP_s$, $\LCP_s$, and an ordered binary search tree of all consecutive intervals in $\LCP_s$, the largest interval of $\SA_s$ that contains all the suffixes of $s$ that share the same prefix of $p$ can be found in $O(\log n)$-time and $O(n)$-space. \end{corollary}

%Lastly, $LCP_s$ is used to determine at each iteration whether a substring is repeated in $s$.  Let $s_{i} \ldots s_n$ be the substring in $s$ for which $p$ is a prefix, and $\ell$ be its index in $SA_s$. If $LCP_s[\ell+1] \geq |p|$ or $LCP_s[\ell-1] \geq |p|$ then we can conclude that $p$ is repeated in $s$.  This can be accomplished in constant time.
   
Theorem \ref{runtime} gives the space and time complexity of our construction algorithm.  See the appendix for the proof of this result. Informally, the time complexity can be seen by first noting that at each iteration of the algorithm the maximal sequence landscape either ascends by one after a number of descents (possible zero) or it is undefined after a number of nonzero descents, and each of these ascents or descents require $O(\log n)$-time.  Note that in (a) the maximal sequence landscape is ascending, and in (b) the maximal sequence landscape is descending, and the frequency of a substring in $s$ can be determined in constant time using $\LCP_s$.   Second, since each time it ascends one character from $s$ is processed and the number of ascents equals the number of descents, the total number of ascents and descents is $2n$. Therefore, since the data structures are constructed in $O(n)$-time, and since there are at most $2n$ ascents or descents which take $O(\log n)$-time, the running time of the algorithm is $O(n + n \log n)$.
 
\begin{theorem} \label{runtime}
The maximal sequence landscape of string $s$ of size $n$ can be built in $O(n + n \log n)$-time and $O(n)$-space. 
\end{theorem}

%% file: approach_readassignment.tex
\subsection{Assignment of $k$-mer Sizes to Reads} \label{sec:alignment} 

We assign values of $k$ to the reads using the maximal sequence landscape constructed for the reference genome by first aligning the reads to the reference genome using BWA (version 0.7.4) \cite{li2009fast} in paired-end mode.  We consider all forward and reverse alignments of every read.  Let $p$ be the position in the reference genome where a read of length $\ell$ aligns, and let $k^*$ be the maximum of $\{\MSL[p] + 1, \MSL[p + 1] + 1, \ldots, \MSL[p + \ell] + 1\}$, where $\MSL$ is the maximal sequence landscape array that contains the height of the maximal sequence landscape at each position. We compute $k^*$ for each forward alignment and let $K^*$ be the set of all these values.  The optimal value of $k$ for the (forward) read is equal to the maximum value in $K^*$. We follow the same procedure for the reverse alignments with the exception that we compute the reverse complement of the read. Thus, the optimal values of $k$ can be computed in linear time in the length of the read. 

If the computed $k$-mer size (maximum of all maximal sequence landscape heights of all aligned nucleotides) is larger than the read length, then a default value ($k = 77$ is the default) is used instead. Unaligned reads are also assigned a default $k$-mer value ($k = 55$ is the default). % If there exists at least one forward alignment then the read and its corresponding $k$-mer size are used for the de Brujin graph construction in HyDA-Vista.  Similarly, if there exists at least one reverse alignment then the reverse complement of the read and its corresponding $k$-mer size are also used.  

% We assign a value of $k$ to all the reads, then the reads are unaligned read $U$ into $k-$mers incrementally by starting from the set of $k-$mer obtained by the guide of the maximal landscape and then for each position of $U$ we search for the smallest $k-$mer that is not already in our pool of $k-$mers, once such a value is found, we add the $k-$mer to our pool of $k-$mers and repeat the process for the next position of $U$. And once we shredded the read into appropriate $k-$mers. we repeat the same process for other unaligned reads. 

%% file: approach_assembly.tex
\subsection{The de Bruijn Graphs} \label{sec:graph}

Let $R=\{r_1,\ldots, r_N\}$ denote the set of reads. We also denote the $k$-mer size assigned to $r_i$ in the previous section by $K(r_i)$. In the first step of constructing the assembly de Bruijn graphs, we partition $R$ into $R_k := \{ r \in R\ |\ K(r) = k\}$, in which $k$ ranges from $k_{\min} = \min_{r \in R} K(r)$ to $k_{\max} = \max_{r \in R} K(r)$. The HyDA-Vista assembly procedure, shown in Algorithm \ref{alg:assembly}, iteratively builds de Bruijn graphs $G_{k_{\min}}, \ldots, G_{k_{\max}}$ with $k=k_{\min},\ldots, k_{\max}$ respectively and obtains $\mathcal{A}_{k_{\min}}, \ldots, \mathcal{A}_{k_{\max}}$ assembly contig sequences after iterative graph condensation and error removal. Each $G_k$ is constructed from the reads whose assigned $k$-mer size is not more than $k$ and the contigs resulting from $G_{k-1}$ constructed in the previous iteration,
\begin{equation}
\bigcup_{j=k_{\min}}^{k} R_j \cup \mathcal{A}_{k-1}.
\end{equation}

The rationale behind this idea is that those reads that have an assigned $k$-mer size not more than $k$ should  ideally not create any repeats when they are assembled with the $k$-mer size $k$. The iterative inclusion of contigs from previous rounds, first introduced in IDBA \cite{peng2010idba} and later adopted by others \cite{bankevich2012spades}, is an idea that has already shown merit in improving assembly quality. In Algorithm \ref{alg:assembly}, \textsc{HyDA} is a function that accepts a set of input sequences and an integer $k$,  and returns a set of contigs which are obtained from assembling the input sequences with a $k$ de Bruijn graph. 

\begin{algorithm}[h!]
\caption{Construction of the de Bruijn graphs}\label{alg:assembly}
\begin{algorithmic}[1]
\Function{HyDA-Vista}{$R, K$} 
\State $k_{\min} \gets \min_{r \in R} K(r)$, $k_{\max} \gets \max_{r \in R} K(r)$ 
\ForAll{$k_{\min} \leq k \leq k_{\max}$}
	\State $R_k \gets \emptyset$
\EndFor
\ForAll{$r \in R$}
	\State $k \gets K(r)$
	\State $R_k \gets R_k \cup \{r\}$
\EndFor
\State $R' \gets \emptyset$
\State $\mathcal{A}_{k_{\min}-1} \gets \emptyset$ \Comment{assembly contigs}
\For{$k := k_{\min}$ {\bf to} $k_{\max}$}
	\State $R' \gets R' \cup R_k$
	\State $\mathcal{A}_k \gets$ \Call{HyDA}{$R' \cup \mathcal{A}_{k-1}, k$} \Comment{contigs resulting from assembly with HyDA}
\EndFor
\State \Return $\mathcal{A}_{k_{\max}}$
\EndFunction
\end{algorithmic}
\end{algorithm}

%% file: results.tex
\section{Results} \label{sec:results} 

\subsection{Improved Efficiency Due to Maximal Sequence Landscape}

HyDA-Vista uses the maximal landscape to break the reads into groups by assigning each a value of $k$. It then uses these groups to build the graph iteratively. This is in contrast to other methods that also iteratively build of the graph; SPAdes \cite{bankevich2012spades} uses \emph{all} the reads at \emph{each} iteration, and IDBA  \cite{peng2010idba} uses a more complicated approach to remove some subset of reads at each iteration.  Thus, one of the main advantages of using the maximal sequence landscape is that it increases the efficiency of building the assembly graph iteratively without degrading assembly quality (see the next subsection for a comparison of the different assemblers).  To demonstrate this efficiency experimentally we ran HyDA-Vista with and without the maximal sequence landscape on multicell {\em E.~coli} (substr.  K-12) Illumina data and the  {\em E.~coli} (substr.  K-12) reference genome.  See Subsection \ref{ecoli} for a description of this dataset.  Without the maximal sequence landscape the assembly took 1,414  minutes, and with the maximal sequence landscape the assembly took 822 minutes with 42 number of minutes for building the maximal sequence landscape and assigning the values of $k$ to the reads. 

\subsection{Comparison Between Competing Assemblers and HyDA-Vista on {\em E.Coli}} \label{ecoli}

The first data set consists of approximately 27 million paired-end 100 bp reads from multicell {\em E.~coli} (substr.  K-12), generated by the Illumina Genome Analayzer (GA) IIx platform. It was obtained from the NCBI Short Read Archive (accession ERA000206, EMBL-EBI Sequence Read Archive). To assess assembly quality, we aligned the reads to the {\em E.~coli} reference genome (substr.  K-12) using BWA (version 0.5.9) \cite{li2009fast} with default parameters.  We call a read {\em mapped} if BWA outputs an alignment for it and {\em unmapped} otherwise.  Analysis of the alignments revealed that 98\% of the reads mapped to the reference genome, representing an average depth of approximately $600\times$;  An analysis using BLAST against known contaminants revealed that the unmapped reads are attributed to minor contamination of the sample \cite{Chitsaz11}.  All reads were error corrected using BayesHammer \cite{bayeshammer} with default parameters. 

KmerGenie \cite{kmergenie13} predicted $41$ to be the optimal $k$-mer value for this dataset.  Therefore, for the assemblers that require a single value of $k$ to be specified (SOAPdenovo, ABySS, HyDA) we  used $k=41$.  HyDA and HyDA-Vista were ran with a cut off of five. All other parameters of SOAPdenovo and ABySS were kept at their default.   SPAdes and IDBA were run with their default parameters in single-end mode. Since the input reads were corrected prior to assembly, the reported data for SPAdes is from the ``only assembly'' stage. IDBA was run with and without error correction, yielding the same statistics as expected.
\begin{table*}[ht!]
\begin{small}

\begin{center}
\caption{The performance comparison between major assembly tools and HyDA-Vista on the error-corrected standard multicell \emph{E.~coli} dataset (6.2 Gbps, 28 million reads, 100bp, treated as single-end) using QUAST in default mode \cite{Gurevich13}. All statistics are based on contigs no shorter than 500 bp. Since there are not (QUAST-defined) misassemblies in any of the assemblies, the length statistics are based on \emph{correct} contigs. NGA50 (NA50) is a (QUAST-corrected) contig size the contigs larger than which cover half of the \emph{genome} (assembly) size \cite{Gurevich13,Earl11}. Total is sum of the length of all contigs. MA is the number of misassemblies. GF is the genome fraction percentage, which is the fraction of genome bases that are covered by the assembly. Unaligned is the total length of all of the contigs that could not be aligned to the reference. }
\begin{tabular}{|c|c|c|c|c|c|c|c|c|c|}
\hline
\textbf{Assembler} 	&\textbf{NGA50}	& \textbf{NA50}	& \textbf{Largest (bp)}	& \textbf{Total (bp) }	&\textbf{MA}	&	\textbf{GF (\%)} & \textbf{Unaligned (bp) }\\ \hline
SOAPdenovo			& 32,032		& 35,343		& 101,201				& 4,304,232			& 3			& 95.2			& 3,421 	\\ \hline
ABySS 				& 	31,237	& 	32,987	& 110,012	& 4,530,701	&	0	& 	97.56	& 	0\\ \hline
SPAdes				& 60,338  	& 60,768 	& 173,976 	& 4,545,775	& 0 & 97.8 	& 3,001 	 \\ \hline
IDBA					& 57,826 	 & 58,549 & 173,964 	& 4,538,426 & 0 & 97.7 	& 2,349	 \\ \hline
HyDA				& 36,292  	& 39,069	& 123,771 	& 4,524,075 	& 0 & 97.4 	& 0 	 \\ \hline
{\bf HyDA-Vista} & {\bf 82,838}  & {\bf 94,910} & {\bf 204,602} & {\bf 4,544,286} & {\bf 0 } & {\bf 97.9} 		&{\bf 0 }	\\ \hline
\end{tabular}
\label{tab:ecolistd}
\end{center}

\end{small}
\end{table*}
\iffalse
\begin{table*}[ht!]
\begin{center}
\caption{\footnotesize{The performance comparison between major assembly tools and HyDA-Vista on simulated {\em Arabidopsis} dataset (0.92 Gbps, 928,630 reads, 100bp, treated as single-end) using QUAST in default mode \cite{Gurevich13}. All statistics are based on contigs no shorter than 500 bp. NGA50, NA50, MA, GF, and ``Unaligned'' are the same as in Table 2.  }}
\label{tab:arab}
\begin{tabular}{|c|c|c|c|c|c|c|c|c|c|}
\hline
\textbf{Assembler} 	&\textbf{NGA50}	& \textbf{NA50}	& \textbf{Largest (bp)}	& \textbf{Total (bp) }	&\textbf{MA}	& \textbf{GF (\%)} & \textbf{Unaligned (bp) }\\ \hline
SOAPdenovo			& 32,121	& 34,002	& 92,023	&  4,523,121	& 0	& 97.5	& 	3.423 \\ \hline
ABySS				&	-		&	2,218	& 	13,737	& 	1,551,003	&	31	&	8.392	&	 408\\ \hline
IDBA				& 7,236 	 & 8,370 	& 62,687 	& 16,512,368  & 2 	& 88.65 & 	15 	 \\ \hline

{\bf HyDA-Vista} 	    	& {\bf 21,065}  	& {\bf 23,354} 	& {\bf 127,990} 			& {\bf 17,085,755 } 	& {\bf 19 }		& {\bf 89.43} 		&	{\bf 5,091 }	     \\ \hline
\end{tabular}
\end{center}
\end{table*}

\fi

Table \ref{tab:ecolistd} gives the standard assembly statistics of all the assemblies.  All statistics in Table \ref{tab:ecolistd} were computed by QUAST in default mode \cite{Gurevich13}.  The results demonstrate that HyDA-Vista achieves the best assembly prior to repeat resolution or scaffolding.   Note that upon determination of $k$-mer sizes, all assemblers were run in single-end mode, i.e., ignoring the pairing information, to study only the effect of our $k$ assignment on contigs. HyDA provides a skeletal de Bruijn graph implementation on which other technologies can be developed.  HyDA alone does not compete with some of the state of the art assemblers such as SPAdes and IDBA that employ multiple sophisticated technologies. However, empowered by only the maximum landscape information without any other sophisticated technology particularly pairing information, HyDA-Vista increases the N50 and NG50 more than twice (in comparison to HyDA) and outperforms the competing assemblers in all measures (the NG50 of 36 kbps obtained by HyDA increases to 82 kbps by HyDA-Vista).   
\iffalse
\subsection{Comparison Between Competing Assemblers and HyDA-Vista on Long Read Sequence Data}

We downloaded the latest version of the Arabidopsis (Col) reference genome from the The Arabidopsis Information Resource (TAIR) website\footnote{https://www.arabidopsis.org} and simulated long (PacBio) error-corrected reads using Chromosome 4 of this reference genome using a custom script.   The script can be downloaded from the HyDA-Vista website.  To the best of our knowledge SPAdes is unable to assemble long reads without short read sequence data and therefore, we compared HyDA-Vista only against HyDA, ABySS, IDBA, and SOAPdenovo. Again, we ran KmerGenie \cite{kmergenie13} to predictthe optimal $k$-mer value for this dataset.  

Table \ref{tab:arab} gives the standard assembly statistics of all the assemblies. Our results demonstrate that HyDA-Vista achieves assembly results that are almost 1.5 times better than other competing assembler, e.g., the NGA50 of HyDA-Vista is 82,838 whereas the best other assembly (IDBA) has an NGA50 of 57,826.  Furthermore, HyDA and HyDA-Vista are the only assembler that have no unaligned regions.   Hence, this experiment demonstrates that for long-reads HyDA-Vista surpasses any competing de Bruijn graph assembler.  Presently long reads have not become as prevalent as short read sequence data, however, in the future they will likely become increasingly common.
\fi

%% file: conclusion.tex
\section{Conclusion} \label{sec:conclusion}

We demonstrated that HyDA-Vista achieves superior performance with respect to standard assembly statistics for ecoli genome before repeat resolution and scaffolding. A crucial aspect of our method is the construction of the maximal sequence landscape for the  phylogenetically-close genome, which allows for the optimal $k$-mer value to be computed for each read.  The maximal sequence landscape requires that a pair of substrings be an exact match in order for them to be considered to be the same repetition.  An area that warrants future investigation is determining if there is an efficient algorithm for computing the maximal sequence landscape for inexact matches, i.e., the maximal sequence landscape with the condition that two substrings $x$ and $y$ are considered to represent the same repetition (mountain) if the Hamming distance or edit distance between the two is at most $d$, where $d$ is a parameter in the problem.   Another open problem is determining whether the maximal sequence landscape could be constructed with only a suffix array.  Since all the data structures (including the suffix array) are constructed in $O(n)$-time and $O(n)$-space the order notation of the running time of such a maximal sequence landscape construction algorithm would likely not improve the running time of the existing algorithm by more than a small constant factor.  However, the removal of the auxiliary data structures may simplify the algorithm and would be of theoretical interest.

\section*{Acknowledgement}
The authors would like to thank Ross McConnell from Colorado State University for many insightful discussions and suggestions. 

%\paragraph{Funding\textcolon} SBSR and CB were funded by the Colorado Clinical and Translational Sciences Institute which is funded by National Institutes of Health (NIH-NCATS,UL1TR001082, TL1TR001081, KL2TR001080).  